\documentclass[aps,pra,twocolumn,floatfix,superscriptaddress]{revtex4-2}

\usepackage{mathrsfs}
\usepackage{graphicx}
\usepackage[english]{babel}
\usepackage{amsmath}
\usepackage{amssymb}

\usepackage{relsize}
\usepackage{CJK}
\usepackage{dsfont}
\usepackage{bm}

\newcommand{\me}{\mathrm{e}}
\newcommand{\mi}{\mathrm{i}}

\newcommand{\dif}{\mathrm{d}}

\newcommand\px{\mathrel{/\mkern-5mu/}}
\allowdisplaybreaks
\begin{document}

\title{Generalizations of Berry phase and differentiation of purified state and thermal vacuum of mixed states}

\author{Xu-Yang Hou}
\affiliation{School of Physics, Southeast University, Jiulonghu Campus, Nanjing 211189, China}

\author{Zi-Wen Huang}
\affiliation{School of Physics, Southeast University, Jiulonghu Campus, Nanjing 211189, China}
\altaffiliation{Current address: Fermi National Accelerator Laboratory (FNAL), Batavia, IL, 60510, USA}
\author{Zheng Zhou}
\affiliation{School of Physics, Southeast University, Jiulonghu Campus, Nanjing 211189, China}
\author{Xin Wang}
\affiliation{School of Physics, Southeast University, Jiulonghu Campus, Nanjing 211189, China}
\author{Hao Guo}
\email{guohao.ph@seu.edu.cn}
\affiliation{School of Physics, Southeast University, Jiulonghu Campus, Nanjing 211189, China}

\author{Chih-Chun Chien}
\email{cchien5@ucmerced.edu}
\affiliation{Department of physics, University of California, Merced, CA 95343, USA}

\begin{abstract}
Two representations of mixed states by state-vectors, known as purified state and thermal vacuum, have been realized on quantum computers. While the two representations look similar, they differ by a partial transposition in the ancilla space. While ordinary observables cannot discern the two representations, we generalize the Berry phase of pure quantum states to mixed states and construct two geometric phases that can reflect the partial transposition. By generalizing the adiabatic condition, we construct the thermal Berry phase, whose values from the two representations can be different, However, the thermal Berry phase may contain non-geometrical contributions. Alternatively, we generalize the parallel-transport condition to include the system and ancilla and show the dynamical phase is excluded under parallel transport. The geometrical phase accumulated in parallel transport is the generalized Berry phase, which may or may not differentiate a purified state from a thermal vacuum depending on the protocol. The generalizations of the Berry phase to mixed states may be realized and measured on quantum computers via the two representations to reveal the rich physics of finite-temperature quantum systems.
\end{abstract}
\maketitle

\section{Introduction}
The Berry phase~\cite{Berry84} has played significant roles in many aspects of physics, ranging from atoms to molecules to condensed-matter systems~\cite{Simon83,BerryZee84,Bohm03,Vanderbilt_book,Cohen19,KaneRMP,ZhangSCRMP,ChiuRMP}. As pointed out in Ref.~\cite{Simon83}, the Berry phase has a profound geometrical origin because an adiabatic and cyclic process of a quantum state is mathematically equivalent to parallel-transporting it along a loop, which connects to the concept of holonomy in geometry~\cite{Bohm03,Nakahara}. Hence, the Berry phase bridges physics and geometry, making it extremely important in the understanding of topological phenomena, such as integer quantum Hall effect, topological insulators and superconductors, and others~\cite{TKNN,Haldane,KaneRMP,ZhangSCRMP,MooreN,KaneMele,KaneMele2,BernevigPRL,MoorePRB,FuLPRL,Bernevigbook,ChiuRMP}.

The description of the Berry phase relies on the properties of a pure state of a quantum systems at zero temperature. Meanwhile, mixed quantum states, including thermal state at finite temperatures, are more common. Therefore, mixed-state generalizations of the Berry phase have been an important task. Uhlmann made a breakthrough by constructing the Uhlmann connection for exploring the topology of finite-temperature systems~\cite{Uhlmann86,Uhlmann89,Uhlmann91,Uhlmann96}. As the Berry holonomy arises from parallel-transport of a state-vector along a closed path, the Uhlmann holonomy is generated by parallel-transporting the amplitude of a density matrix. defined by $W=\sqrt{\rho}U$. Here the amplitude $W$ is the mixed-state counterpart of the wavefunction, and $U$ is a phase factor. A geometrical phase is deduced from the initial and final amplitudes. However, Uhlmann's definition of parallel-transport is rather abstract and may involve nonunitary processes~\cite{OurPRB20}, complicating a direct and clear physical interpretation. Moreover, the fiber bundle built upon Uhlmann's formalism is trivial~\cite{DiehlPRB15}, which severely restricts its applications in physical systems.

Purification of a mixed state leads to purified state, a state-vector equivalent to the amplitude of a density matrix. The lack of a one-to-one correspondence between the density matrix and its purified states gives rise to a phase factor, similar to the phase of a wavefunction. In a branch of quantum field theory called thermal field theory~\cite{TFD_book,TFD_book2,Vitiello_book,Cottrell19}, there is a similar structure for describing the thermal-equilibrium state of a system by constructing the corresponding thermal vacuum (or thermofield double state) by duplicating the system state as an ancilla and forming a composite state. It plays a crucial role in the formalism of traversable wormholes induced by the holographic correspondence between a quantum field theory and a gravitational theory of one higher dimensions~\cite{GaoJHEP17,SusskindPRD18}. Importantly, purified states of a two-level system has been demonstrated on the IBM quantum computer~\cite{npj18} while the thermal vacuum of a transverse-field Ising model in its approximate form has been realized on a trapped-ion quantum computer~\cite{TFDPNAS20}.

Despite the superficial similarity, a major difference between a thermal vacuum and a purified state is a partial transposition of the ancilla to ensure the Hilbert-Schmidt product is well defined.
In quantum information theory, a partial transposition is closely related to entanglement of mixed states~\cite{QI_book}. Importantly, partial transpositions of composite systems have been approximately realized in experiments by utilizing structural physical approximations in suitable quantum computing platforms~\cite{SPAPRA11,SPAPRL11,SPAPRA12}. Although ordinary observables cannot discern the partial transposition between the purified state and thermal vacuum, here we will show that at least two types of generalizations of the Berry phase to mixed states are capable of differentiating the two representations of finite-temperature systems.

Among many attempts to generalize the Berry phase or related geometric concepts to mixed states~\cite{GPMQS1,GPMQS2,GPMQS3,GPMQS4,GPMQS5,Andersson13,DiehlPRX17,WangSR19}, a frequently mentioned approach was proposed in Ref.~\cite{GPMQS1}. Instead of decomposing the density matrix to obtain a matrix-valued phase factor, a geometrical phase is directly assigned to a mixed state after parallel transport by an analogue of the optical process of the Mach-Zehnder interferometer. Hence, the geometrical phase generated in this way is often referred to as the interferometric phase. This approach can also be cast into a formalism by rewriting a mixed state as a purified state. The interferometric phase has been generalized to non-unitary processes~\cite{PhysRevA.67.020101,PhysRevLett.90.160402,Faria_2003,PhysRevLett.93.080405,GPM06}, but the transformations are still on the system only. Moreover, it is essentially different from Uhlmann's theory since the conceptual structure of holonomy is not incorporated.

We will first derive a mixed-state generalization of the parallel-transport condition for generalizing the Berry phase without invoking holonomy. This approach unifies the necessary condition for both the interferometric phase and Uhlmann phase (in a weak version). Two ways to implement the parallel-transport condition based on how the system of interest undergoes adiabatic evolution will be introduced, and they lead to different generalizations of the Berry phase. We will name one thermal Berry phase and the other generalized Berry phase. Importantly, the partial transposition of the ancilla between a purified state and thermal vacuum will be shown to produces observable geometrical effects in both thermal Berry phase and generalized Berry phase. Through explicit examples, the two generalized phases are shown to differentiate the two finite-temperature representations, a task beyond the capability of the conventional interferometric phase or Uhlmann phase.

The rest of the paper is organized as follows. Sec. \ref{SecBerry} summarizes the Berry phase in a geometrical framework with an introduction of the parallel-transport condition for pure quantum states. In Sec. \ref{Sec3}, we review the representations of mixed states via purified states and thermal vacua and then explain the difference of the partial transposition of the ancilla.
In Sec. \ref{Sec4}, we introduce the thermal Berry phase via generalized adiabatic processes. While the thermal Berry phase can differentiate a purified state from a thermal vacuum, it may contain non-geometrical contributions.
In Sec. \ref{Sec5}, we generalize the parallel-transport condition to involve the system and ancilla and derive the general Berry phase according to the generalized condition. While the generalized Berry phase only carries geometrical information, its ability of differentiating a purified state from a thermal vacuum depends on the setup and protocol. We present examples of the thermal and generalized Berry phases. Sec. \ref{Sec6} concludes our study. Some details and derivations are given in the Appendix.

\section{Berry Phase and Parallel Transport}\label{SecBerry}
We first give a brief review of the Berry phase of pure states by quoting some results~\cite{Berry84} but from the viewpoint of parallel transport~\cite{Uhlmann86,ourPRB20b}, which will be a starting point for generalizations to mixed quantum states.
For simplicity, we only focus on systems with no degenerate energy levels.
Consider a quantum system described by the Hamiltonian $\hat{H}(\mathbf{R})$, where $\mathbf{R}=(R_1,R_2,\cdots, R_k)$ is a collection of parameters. If the system undergoes an adiabatic cyclic process denoted by a closed curve $\mathbf{R}(t)$ in the parameter space, then the $n$th energy level obtains the geometrical phase known as the Berry phase:
\begin{align}\label{Bp1}
\theta_{n}=\mi\int_0^\tau \dif t\langle n\left(\mathbf{R}(t)\right)|\frac{\dif}{\dif t}|n\left(\mathbf{R}(t)\right)\rangle.
\end{align}
Here $t$ is a parameter not necessarily being the time, and $\tau$ is the ``duration'' of this process. This formalism can be reformulated by the language of fiber bundles~\cite{Simon83,TDMPRB15,OurPRB20}, which lays the foundation for the generalization to the Uhlmann phase of mixed states~\cite{Uhlmann86,Uhlmann89,Uhlmann91}. Here we adopt a simpler geometric illustration based on parallel transport of quantum states.

To measure the similarity between two pure quantum states, the quantum fidelity~\cite{WatrousBook} is introduced to quantify the overlap between the states of interest. Moreover, Uhlmann defined the parallelity between quantum states~\cite{Uhlmann86}: Two pure states $|\psi_1\rangle$ and $|\psi_2\rangle$ are said to be parallel to each other (or in phase) if
\begin{align}
\langle \psi_1|\psi_2\rangle=\langle \psi_2|\psi_1\rangle>0,
\end{align}
i.e., the quantum fidelity between them is a positive number and $\arg\langle \psi_1|\psi_2\rangle=0$.
This may be thought of as an antonym of the orthogonal condition between two states,
$\langle \psi_1|\psi_2\rangle=0$,
which means $|\psi_{1,2}\rangle$ have minimal similarity since they share no common components.

However, parallelity does not define an equivalence relation between quantum states since it possesses only reflexivity
and symmetry but lacks transitivity. It can be shown that $|\psi_1\rangle  \px |\psi_2\rangle$ and $|\psi_2\rangle  \px |\psi_3\rangle$ do not necessarily imply $|\psi_1\rangle  \px |\psi_3\rangle$. Now consider a physical process in which the state evolves as $|\psi(t)\rangle$. During parallel transport, the parallel condition is preserved as much as possible. For an infinitesimal change of the parameter $t$, we have
\begin{align}\label{pc1}
\langle \psi(t)|\psi(t+\dif t)\rangle=\langle \psi (t+\dif t)|\psi (t)\rangle>0.
\end{align}
However, since this condition is not transitive, the final state $|\psi(\tau)\rangle$ may not be parallel to the initial state $|\psi(0)\rangle$ even if the process is cyclic. Thus, the quantum fidelity between them may be a complex number whose argument gives the geometric phase generated in this process. 
Note $\langle\psi(t)|\psi(t+\dif t)\rangle\approx1$ since $\dif t$ is small. Taking the Taylor expansion of $|\psi(t+\dif t)\rangle$, the left hand side of Eq. (\ref{pc1}) becomes
$\langle\psi(t)|\psi(t+\dif t)\rangle
\approx 1+\langle\psi(t)|\frac{\dif}{\dif t}|\psi(t)\rangle\dif t $.
Here $\langle\psi(t)|\frac{\dif}{\dif t}|\psi(t)\rangle$ is a purely imaginary number, but $\langle\psi(t)|\psi(t+\dif t)\rangle$ is real according to the parallel condition.  Hence, to ensure the parallel-transport condition, we must have
\begin{align}\label{PXT}
\langle\psi(t)|\frac{\dif}{\dif t}|\psi(t)\rangle=0.
\end{align}
Note the left-hand-side is purely imaginary, so it is equivalent to
$\text{Im}\langle\psi(t)|\frac{\dif}{\dif t}|\psi(t)\rangle=0$.
This is the parallel-transport condition for pure states. Physically, it means the instantaneous change of a state must be perpendicular to itself at any time to preserve the instantaneous parallel condition.

It can be shown that although the pure dynamical evolution solely governed by the Hamiltonian qualifies as an adiabatic process, it does not fulfill the parallel-transport condition.
For simplicity, suppose $|\psi(0)\rangle=|n\rangle$, i.e., the $n$th energy level. The parameter $t$ is then identified as the time, and the dynamic time evolution leads to
\begin{align}\label{dp1}
|\psi(t)\rangle=\me^{-\frac{\mi}{\hbar}\mathlarger{\int}_0^t E_n(t')\dif t'}|n\rangle,
\end{align}
which is adiabatic if no level crossing occurs. The phase accumulated is known as the dynamical phase. Then,
\begin{align}\label{PXT1}
\langle\psi(t)|\frac{\dif}{\dif t}|\psi(t)\rangle=-\frac{\mi}{\hbar}E_n,
\end{align}
which is nonzero in general. Interestingly, $|\psi(t)\rangle$ is physically equivalent to but not parallel to the initial state. To follow parallel transport, this process must carry some extra geometric information, such as the topological properties of $\mathbf{R}(t)$~\cite{Nakahara} if the process is induced by $\mathbf{R}(t)$. Nevertheless, to ensure the validity of the parallel-transport condition, one should remove the dynamical phase from Eq. (\ref{dp1}) since it contributes to Eq. (\ref{PXT1}).

Here is a geometric description of the Berry phase: Consider a quantum system initially in the state $|\psi(0)\rangle\equiv|n(0)\rangle$. During a process where the state is parallel-transported along a curve $C(t):=\mathbf{R}(t)$ in the parameter space, the parallel transport preserves the similarity between the initial and successive states as much as possible, but the parallelity may still be lost.
We assume $|\psi(t)\rangle=\me^{\mi\theta_{n}}|n(\mathbf{R(t)})\rangle$, where $\theta_{n}$ is the phase accumulated in the process.
For a cyclic process, $C$ is closed, and the system returns to a state that is equivalent to the initial state. However, they are not necessarily parallel. The similarity is measured by the fidelity between the initial and final states, $\langle\psi(0)|\psi(\tau)\rangle$, and the Berry phase is the argument of the fidelity:
\begin{align}\label{Bp0}
\theta_{n}=\arg\langle\psi(0)|\psi(\tau)\rangle.
\end{align}
Substituting $|\psi(t)\rangle$ into the parallel-transport condition (\ref{PXT}),
we get
\begin{align}\label{SBp}
\mi\frac{\dif\theta_{n}}{\dif t}+\langle n(\mathbf{R(t)})|\frac{\dif }{\dif t}|n(\mathbf{R(t)})\rangle=0.
\end{align}
Solving this equation, the Berry phase is
exactly the same as Eq. (\ref{Bp1}). Moreover, the dynamical phase has been excluded by the parallel-transport condition.  Therefore, $\theta_n$ is a measure of the loss of parallelity during parallel transport, so it is endowed with a geometrical meaning.

\section{State-vector representations of density matrix}\label{Sec3}
Generalizations of the parallelity, parallel transport, and geometrical phase to mixed quantum states are possible by introducing an ancilla to cast a mixed state into a state-vector of the augmented system. In quantum statistics, a mixed state is usually represented by a density matrix, which is a Hermitian operator and does not carry any phase information on its own. In the literature, two state-vector representations of density matrix have been extensively studied. Here we briefly review and compare them.

\subsection{Purification of Density Matrices}
The concept of purification has been applied broadly in quantum information~\cite{QCQI_book,WatrousBook,QI_book}. To investigate geometrical phase associated with purification, we follow Uhlmann's approach~\cite{Uhlmann86} that a full-rank density matrix $\rho$ can be uniquely decomposed as
\begin{align}\label{dofr}
\rho=WW^\dagger,
\end{align}
where $W$ is called the amplitude or purification of the mixed state. Conversely, a full-rank matrix $W$ also has a unique polar decomposition $W=\sqrt{\rho}U$, where the unitary matrix $U$ may be recognized as the ``phase factor'' of $\rho$. While the phase factor of a pure state is in general not unique, the phase factor from purification is also not uniquely given. If the dimension of $\rho$ is $N$,
there is a U$(N)$ degrees of freedom in the choice of the amplitude because both $W$ and $WV$ are amplitudes of the same density matrix $\rho$ with an arbitrary unitary operator $V\in$ U$(N)$. Effectively, the amplitudes play the role of wavefunctions (state vectors), and they
form a Hilbert space $H_W$ in which a scalar product, called the Hilbert-Schmidt product, is defined as
\begin{align}\label{HSip}
(W_1,W_2):=\textrm{Tr}(W^\dagger_1W_2).
\end{align}

Assuming the space spanned by the eigenvectors $|n\rangle$ of $\rho$ is $\mathcal{H}$, $\rho$ can be expressed as
$\rho=\sum_n\lambda_n|n\rangle\langle n|$. The amplitude has the form
$W=\sum_n\sqrt{\lambda_n}|n\rangle\langle n|U $.
Each $W$ corresponds to a wavefunction of the form
\begin{align}\label{w2}
W\leftrightarrow|W\rangle=\sum_n\sqrt{\lambda_n}|n\rangle\otimes U^T|n\rangle,
\end{align}
where $|W\rangle$ is called the purified state of $\rho$. Here $U^T$ is the transpose of $U$ and only acts on the second Hilbert space usually called the ancilla.
In this way, the density matrix is effectively represented by an entangled pure state consisting of the original system and the ancilla. This form allows simulations of a mixed state on a (quantum) computer via its purified states~\cite{npj18}.

Although $W\leftrightarrow|W\rangle$ is an isomorphism between $H_W$ and $\mathcal{H}\otimes\mathcal{H}$, the inner product in $\mathcal{H}\otimes\mathcal{H}$ in its naive form does not reproduce the Hilbert-Schmidt product in $H_W$.
This actually raises a subtle requirement to the inner product involving $|W\rangle$.
To make the point explicit, we suppose $W_{1,2}$ both purify $\rho$, so $W_{1,2}=\sqrt{\rho}U_{1,2}$ or $|W_{1,2}\rangle=\sum_n\sqrt{\lambda_n}|n\rangle\otimes U^T_{1,2}|n\rangle$.
The Hilbert-Schmidt product is
\begin{align}\label{ipe2}
\textrm{Tr}(W^\dagger_1W_2)=\textrm{Tr}(U^\dagger_1\sqrt{\rho}\sqrt{\rho}U_2)=\textrm{Tr}(\rho U_2U^\dagger_1).
\end{align}
If $\langle W_{1,2}|$ is expressed as the Hermitian conjugate of $|W_{1,2}\rangle$, $\langle W_{1,2}|=\sum_n\sqrt{\lambda_n}\langle n|\otimes\langle n|U^\ast_{1,2}$, then
\begin{align}\label{ipe3}
\langle W_1|W_2\rangle&=\sum_{n,m}\sqrt{\lambda_n\lambda_m}\langle n|m\rangle \langle n|U^\ast_1 U^T_2|m\rangle\notag\\
&=\textrm{Tr}(\rho U^\ast_1 U^T_2),
\end{align}
which differs from Eq. (\ref{ipe2}).
To satisfy the Hilbert-Schmidt inner product
\begin{align}\label{ipe1}
	\langle W_1|W_2\rangle = \textrm{Tr}(W^\dagger_1W_2),
\end{align}
we will forgo the transpose in the second Hilbert space when calculating inner products. Thus,
\begin{align}\label{ipe4}
\langle W_1|W_2\rangle&=\sum_{n,m}\sqrt{\lambda_n\lambda_m}\langle n|m\rangle \langle m|U_2 U^\dagger_1|n\rangle\notag\\
&=\textrm{Tr}(\rho  U_2 U^\dagger_1).
\end{align}
We emphasize that this seemingly minor detail in fact plays an important role when measuring the Uhlmann phase in a quantum simulator by fulfilling a weakened parallel-transport condition~\cite{ourPRA21}. In quantum information theory, the structure is also referred to as the partial transposition of the ancilla and is closely related to the separability of density matrices~\cite{QI_book}.

\subsection{Thermal Vacuum}
The framework of purification can be applied to generic mixed states. In quantum field theory or condensed matter physics, one often deals with finite-temperature phenomena of many-body systems in equilibrium, which are also described by mixed states with thermal distributions. Many finite-temperature field theories have been developed, including the Matsubara formalism for equilibrium or near-equilibrium systems~\cite{Matsubara,AGD_book} and the Keldysh formalism~\cite{Keldysh,Kapusta_book} for non-equilibrium systems. Meanwhile, thermal field theory~\cite{TFD_book,TFD_book2,Vitiello_book} is built through thermal vacuum that also has the form of pure states. As we will show, a thermal vacuum bears many properties of a temperature-dependent purified state of the density matrix. Interestingly, thermal vacuum may also be simulated on quantum computers~\cite{TFDPNAS20}.

For a mixed state in thermal equilibrium described by the canonical ensemble, the density matrix is given by $\rho=\frac{1}{Z}\me^{-\beta \hat{H}}$, where  $Z=\mathrm{Tr}\mathrm{e}^{-\beta \hat{H}}$ is the partition function at inverse temperature $\beta=\frac{1}{k_B T}$. For an observable $\mathcal{O}$, its expectation value is given by
\begin{align}\label{aO}
\langle \mathcal{O}\rangle=\text{Tr}(\mathcal{O}\rho)=\frac{1}{Z}\mathrm{Tr} (\mathrm{e}^{-\beta \hat{H}}\mathcal{O}).
\end{align}
The central idea of thermal field theory is to
replace the statistical average (\ref{aO})
by the expectation value of $\mathcal{O}$ in a formally temperature-dependent pure state, called a thermal vacuum $|0_\beta\rangle$~\cite{QP_book,Vitiello_book}. Explicitly,
\begin{equation}\label{TFD1}
\langle \mathcal{O}\rangle=\langle0_\beta|\mathcal{O}|0_\beta\rangle.
\end{equation}
This can be achieved by doubling the degrees of freedom of the system via
\begin{align}\label{ThV1}
|0_\beta\rangle=\frac{1}{\sqrt{Z}}\sum_n\me^{-\frac{\beta \hat{H}}{2}}|n\rangle\otimes |\tilde{n}\rangle,
\end{align}
where $|\tilde{n}\rangle$ is a duplication of $|n\rangle$ and is characterized as the eigenvector of the ``tilde" system with the ``tilde'' Hamiltonian $\tilde{H}$~\cite{Vitiello_book}.
Note the operator $\mathcal{O}$ only acts on the first Hilbert space spanned by $|n\rangle$ in Eq. (\ref{TFD1}). When compared to Eq. (\ref{w2}), it can be found that $|0_\beta\rangle$ has a similar structure as the purified state $|W\rangle$ since the tilde space is like an ancilla. In fact, $|W\rangle$ also satisfies the definition (\ref{aO}):
\begin{align}\label{aO1}
\langle W|\mathcal{O}|W\rangle
=\sum_{n,m}\sqrt{\lambda_n\lambda_m}\langle m|UU^\dagger|n\rangle\langle n|\mathcal{O}|m\rangle
=\text{Tr}(\mathcal{O}\rho).
\end{align}
Thus, if $\rho=WW^\dagger$ describes a mixed state in thermal equilibrium, $|W\rangle$ qualifies as a type of thermal vacuum. Interestingly, Eq. (\ref{aO1}) is satisfied regardless of the transposition in the ancilla space, indicating that expectation values are not good indicators of the difference between a purified state and thermal vacuum.
However, a thermal vacuum in general may not be a purified state since the inner product of $|0_\beta\rangle$ may not satisfy the Hilbert-Schmidt product. They are thus two ways to construct pure-state representations of mixed states.

For a system in thermal equilibrium, $[\rho, \hat{H}]=0$, so $\rho$ and $\hat{H}$ share the same eigenvectors. Let $\hat{H}|n\rangle=E_n|n\rangle$, then $\lambda_n=\frac{1}{\sqrt{Z}}\me^{-\frac{\beta E_n}{2}}$. Therefore, a typical thermal vacuum is expressed as
\begin{align}\label{ThV2}
|0_\beta\rangle=\frac{1}{\sqrt{Z}}\sum_n\me^{-\frac{\beta E_n}{2}}|n\rangle\otimes \tilde{U}^T|\tilde{n}\rangle \quad (\text{with } n=\tilde{n}),
\end{align}
where the unitary matrix $\tilde{U}^T$ with an extra transpose is included to formally match the general expression (\ref{w2}) of $|W\rangle$ for convenience, and the tilde symbol is used for any object related to the ancilla of a thermal vacuum hereafter.
From here on, we refer to $|W\rangle$ as a purified state and $|0_\beta\rangle$ as a thermal vacuum.
The difference between them is whether a partial transposition is applied to the ancilla.
As a consequence, the inner product in the ancilla space of a purified state is evaluated by following Eq. (\ref{ipe4}) while that of a thermal vacuum is evaluated in a way similar Eq. (\ref{ipe3}) due to the lack of the constraint from the Hilbert-Schmidt product. 
Here we are interested in whether any geometrical effect can be inferred from this subtle difference.

Alternatively, a thermal vacuum can be obtained by a temperature-dependent unitary transformation on the two-mode ground state of $\hat{H}\otimes \tilde{H}$:
\begin{align}\label{UT}
|0_\beta\rangle=U_\beta(|0\rangle\otimes|\tilde{0}\rangle)\equiv U_\beta|0,\tilde{0}\rangle.
\end{align}
Thus, $|0_\beta\rangle$ is formally the ground state of the ``thermal Hamiltonian" $\hat{H}_\beta=U_\beta \hat{H}U_\beta^{-1}$, i.e., $\hat{H}_\beta|0_\beta\rangle=E_0|0_\beta\rangle$. This is why $|0_\beta\rangle$ is called a thermal vacuum: It is the vacuum state of $\hat{H}_\beta$ and is also temperature-dependent. Clearly, $\hat{H}_\beta$ describes a large system when compared to $\hat{H}$ since its ground state already contains all information of the system governed by $\hat{H}$. However, higher energy levels of $\hat{H}_\beta$, i.e., $|n_\beta\rangle=U_\beta|n,\tilde{n}\rangle$, are usually irrelevant to our discussions. The explicit expression of $U_\beta$ depends on the details of the Hamiltonian, and we will show its expression of an exemplary two-level system in out later discussion.

\section{Thermal Berry Phase}\label{Sec4}
An important task is to check if any geometric phase can differentiate the two representations of mixed states that can be realized on quantum computers~\cite{npj18,TFDPNAS20}. We have reviewed the literature on generalizations of the Berry phase~\cite{Uhlmann86,GPMQS1,GPMQS2,GPMQS3,GPMQS4,GPMQS5,Andersson13,DiehlPRX17,WangSR19} but could not find suitable candidates. For example, the theory of the Uhlmann phase explicitly depends on the Hilbert-Schmidt product (\ref{ipe1}), which is not applicable to thermal vacuum by its construction. The interferometric phase~\cite{GPMQS1} and its subsequent developments~\cite{PhysRevA.67.020101,PhysRevLett.90.160402,Faria_2003,PhysRevLett.93.080405,GPM06} are generated by transformations only acting on the system, so it is insensitive to operations on the ancilla and not able to differentiate the two representations as well. In the following, we will construct our first generalization that can differentiate the purified state and thermal vacuum. 

\subsection{Generalization of the parallel-transport condition and Berry phase}
Equipped with the two pure-state representations of mixed states, we are now ready to generalize the concepts of parallel transport and geometrical phase to mixed states. There are several ways to accomplish this, based on the choices of quantum systems on which the parallel-transport condition is imposed. We will demonstrate two generalizations of the Berry phase and compare the results from a purified state and a thermal vacuum.

The first generalization is derived from thermal field theory, but it can be applied to purified states as well. Note $|0_\beta\rangle$ is the ground state of the thermal Hamiltonian $\hat{H}_\beta$. Hence, an extension of the Berry phase starts with a thermal system governed by $\hat{H}_\beta $ undergoing a cyclic adiabatic process along a closed curve $C(t)=\mathbf{R}(t)$ in the parameter space. Let the instantaneous thermal ground-state at $t$ be $|\Psi(t)\rangle=\me^{\mi\theta_{TB}(t)}|0_\beta(\mathbf{R}(t))\rangle$.
We assume that the system is always in thermal equilibrium, so the density matrix and the Hamiltonian have common eigenvectors. Similar to the geometric formalism of the Berry phase, the parallel-transport condition can be obtained by replacing $|\psi(t)\rangle$ in Eq. (\ref{PXT}) by the $t$-dependent thermal vacuum:
\begin{equation}\label{Bptv}
\langle\Psi(t)|\frac{\dif}{\dif t}|\Psi(t)\rangle=0.
\end{equation}
Note $|0_\beta\rangle$ is the ground state of the thermal Hamiltonian $\hat{H}_\beta$, hence the name thermal vacuum. Just like its pure-state counterpart, the parallel-transport condition accordingly eliminates the dynamical phase factor $\me^{-\frac{1}{\hbar}\mathlarger{\int}_0^t\dif t'E_0(\mathbf{R}(t))}$ generated by the time evolution via   
\begin{equation}\label{Dp}
	U_D(t)=\me^{-\frac{\mi}{\hbar}\mathlarger{\int}_0^t\hat{H}_\beta(\mathbf{R}(t))\dif t'}.
\end{equation}
We emphasize that the dynamical evolution (\ref{Dp}) is governed by the thermal Hamiltonian $\hat{H}_\beta$ instead of $\hat{H}\otimes \tilde{H}$, and $\hat{H}_\beta|0_\beta\rangle=E_0|0_\beta\rangle$. Thus, only the single phase factor $\me^{-\frac{1}{\hbar}\mathlarger{\int}_0^t\dif t'E_0(\mathbf{R}(t))}$ of $|0_\beta\rangle$ is removed, and the dynamical phases of the higher two-mode levels $|n\rangle\otimes |\tilde{n}\rangle$ of $\hat{H}\otimes \tilde{H}$ are irrelevant here. 
Substituting $|\Psi(t)\rangle=\me^{\mi\theta_{TB}(t)}|0_\beta(\mathbf{R}(t))\rangle$ into Eq. (\ref{Bptv}) and following Eq. (\ref{SBp}) to solve it, we get
\begin{align}\label{TB}
\theta_{TB}(C)
&=\mi\oint_C\dif t\langle 0_\beta\left(\mathbf{R}(t)\right)|\frac{\dif}{\dif t}|0_\beta\left(\mathbf{R}(t)\right)\rangle.
\end{align}
Since this generalization of the Berry phase is easier to obtain from the thermal-vacuum formalism, we refer to it as the thermal Berry phase hereafter.

In this generalization, it is the ``thermal quantum system'' governed by $\hat{H}_\beta$ that experiences an adiabatic evolution. This may sound artificial since the dependence of $\hat{H}_\beta$ on temperature obscures the physical interpretation of this process. Therefore, the adiabatic evolution and dynamical evolution (\ref{Dp}) of such a thermal quantum system may not occur in a natural way but may be simulated by engineered systems or on quantum computers. Nevertheless, some interesting results will be predicted under this generalization of the Berry phase.

\subsection{Theoretical predictions}
To derive the expression of the thermal Berry phase, one must be careful about whether the representation of the ancilla is for a thermal vacuum or purified state. Here we present two separate discussions.

\subsubsection{Thermal vacuum}
For a thermal vacuum, the state vector in the ancilla follows Eq. (\ref{ThV2}).
Note the whole composite thermal system governed by $\hat{H}_\beta$ is supposed to undergo an artificial adiabatic evolution along $\mathbf{R}(t)$ at temperature $T$ (with $\beta=\frac{1}{k_BT}$). Accordingly, the most general form of a thermal vacuum is written as
\begin{align}\label{ThV3}
|0_\beta(t)\rangle\equiv\frac{1}{\sqrt{Z(t)}}\sum_n\me^{-\frac{\beta E_n(t)}{2}}|n(t)\rangle\otimes \tilde{U}^T(t)|\tilde{n}(t)\rangle,
\end{align}
where $Z(t)\equiv Z(\mathbf{R}(t))$, $E_n(t)\equiv E_n(\mathbf{R}(t))$, $|n(t)\rangle\equiv|n,\mathbf{R}(t)\rangle$, and $\tilde{U}(t)\equiv \tilde{U}(\mathbf{R}(t))$. A straightforward calculation by using Eq. (\ref{TB}) shows
\begin{align}\label{ThV4}
\theta_{TB}&=\mi\int_0^\tau\dif t\bigg\{-\frac{1}{2Z(t)}\sum_n\me^{-\beta E_n(t)}\left(\frac{\dot{Z}(t)}{Z(t)}+\beta \dot{E}_n(t)\right)\notag\\
&+\frac{1}{Z(t)}\sum_n\me^{-\beta E_n(t)}\Big[\langle n(t)|\frac{\dif }{\dif t}|n(t)\rangle+\langle\tilde{n}(t)|\frac{\dif }{\dif t}|\tilde{n}(t)\rangle\notag\\&+\langle\tilde{n}(t)|\tilde{U}^*(t)\dot{\tilde{U}}^T(t)|\tilde{n}(t)\rangle\Big]\bigg\}.
\end{align}
Applying
$\dot{Z}(t)=-\sum_n\beta \dot{E}_n(t)\me^{-\beta E_n(t)}$ and  $\sum_n\me^{-\beta E_n(t)}=Z(t)$,
the first line of the right-hand-side of Eq. (\ref{ThV4}) vanishes. Moreover, since $|\tilde{n}(t)\rangle$ is a copy of $|n(t)\rangle$, we finally get
 \begin{align}\label{ThV4a}
\theta_{TB}&=\mi\int_0^\tau\dif t\sum_n\frac{\me^{-\beta E_n(t)}}{Z(t)}\Big[2\langle n(t)|\frac{\dif }{\dif t}|n(t)\rangle\notag\\&+\langle n(t)|\tilde{U}^*(t)\dot{\tilde{U}}^T(t)|n(t)\rangle\Big],
\end{align}
or equivalently
\begin{align}\label{ThV4b}
\theta_{TB}=2\mi\int_0^\tau\dif t\text{Tr}_t\left[\rho(t)\left(\frac{\dif }{\dif t}+\frac{1}{2}\tilde{U}^*(t)\dot{\tilde{U}}^T(t)\right)\right],
\end{align}
where $\text{Tr}_t$ means the trace is over the instantaneous states $|n(t)\rangle$, $n=1,2,\cdots, N$. Here the factor 2 comes from the contributions of both the system and ancilla.

Let $X$ be the tangent vector of the curve $C(t)$, which is locally expressed as $X(t)=\frac{\dif}{\dif t}$. For each energy level, we introduce the Berry connection
\begin{align}\label{BAn}
A^n_{B}=\mi\langle n(\mathbf{R})|\dif|n(\mathbf{R})\rangle.
\end{align}
Thus, the first term of $\theta_{TB}$ is
$2\int_0^\tau\sum_n\frac{\me^{-\beta E_n(t)}}{Z(t)}A^n_B(X(t))\dif t $,
or equivalently
\begin{align}\label{ThV4c2}
2\oint_C\sum_n\frac{\me^{-\beta E_n(\mathbf{R})}}{Z(\mathbf{R})}\mathbf{A}^n_B(\mathbf{R})\cdot \dif \mathbf{R},
\end{align}
where $\mathbf{A}^n_{B}=\mi\langle n(\mathbf{R})|\nabla|n(\mathbf{R})\rangle$ is the Berry curvature associated with the $n$th level. Note $\mathlarger{\oint}_C\mathbf{A}^n_B(\mathbf{R})\cdot \dif \mathbf{R}$ is actually the original Berry phase that the $n$th level acquires during the evolution. Therefore, this term is the weighted sum of the Berry phase of each level. Similar result can be found in Ref. \cite{GPmPRA05}, which provides a direct comparison to the interferometric phase.

The first term of $\theta_{TB}$ carries geometrical information about the topology of $C(t)$ in the parameter space, which will be elaborated by an explicit example later. However, the role of the second is not so clear. At first look, its expression is not so ``geometrical''. If we introduce the connection 1-form
\begin{align}\label{AU}A_U=\frac{1}{2}\tilde{U}^*\dif \tilde{U}^T=\frac{1}{2}\tilde{U}^*\dif (\tilde{U}^*)^\dagger,
\end{align}
then $\frac{1}{2}\tilde{U}^*(t)\dot{\tilde{U}}^T(t)=A_U(X(t))$ can be understood as an effective gauge potential, and the thermal Berry phase can be formally expressed as
\begin{align}\label{ThV4b2}
&\theta_{TB}=2\mi\int_0^\tau\dif t\text{Tr}_t\left[\rho(t)\nabla^U_X\right]\notag\\
&=2\mi\int_0^\tau\dif t\sum_n\frac{\me^{-\beta E_n(t)}}{Z(t)}\langle n(t)|\left(\frac{\dif }{\dif t}+A_U(X(t))\right)|n(t)\rangle,
\end{align}
where $\nabla^U$ is the covariant derivative induced by the connection $A_U$. Note $A_U$ is a pure gauge, thus the associated curvature is zero and has no geometrical effect accordingly.
Moreover, $A_U$ may not necessarily be a diagonal matrix with respect to the basis $\{|n(t)\rangle\}$. Thus, it may cause transitions between different energy levels. This is not surprising since what undergoes an adiabatic evolution is the quantum system governed by $\hat{H}_\beta$ but not $\hat{H}$. Similarly, the parallel-transport condition (\ref{Bptv}) only excludes the artificial time evolution (\ref{Dp}), which is governed by $\hat{H}_\beta$. Time evolution generated by $\hat{H}$ or $\tilde{H}$ is not forbidden.

For example, if the system is kept unchanged and the ancilla undergoes time evolution generated by a time-independent Hamiltonian $\tilde{H}$, then the only time-dependent part of the thermal vacuum ~\eqref{ThV3} is
\begin{align}\label{ThB1b}
\tilde{U}^T(t)=\left(\begin{array}{cccc}
\me^{-\frac{\mi}{\hbar}E_0t} & 0 & 0 & \cdots \\
0 & \me^{-\frac{\mi}{\hbar}E_1t} & 0 &  \cdots \\
0 & 0 & \me^{-\frac{\mi}{\hbar}E_2t} & \cdots \\
\vdots & \vdots & \vdots & \ddots \end{array}\right).
\end{align}
The thermal Berry phase in this case is actually the thermal dynamical phase generated by $\tilde{H}$:
\begin{align}\label{ThV4d}
\theta_{TB}=\sum_n\frac{\me^{-\beta E_n}}{Z}\frac{E_n}{\hbar}\tau=\frac{\bar{E}\tau}{\hbar},
\end{align}
where $\bar{E}$ is the average energy of the ancilla. Therefore, the thermal Berry phase is more than just a geometrical phase, it carries more information than the original Berry phase since a thermal vacuum has more degrees of freedom than a pure state of the system. For this reason, the thermal Berry phase may not be a simple geometrical phase of mixed states due to the possible inclusion of the dynamical phase.
If we want the thermal Berry phase to be geometrical according to the conventional sense, we can set $\tilde{U}=1$ in Eq. (\ref{ThV3}) to obtain the conventional result.

\subsubsection{Purified state}
After considering the thermal Berry phase of a thermal vacuum, we consider the analogous case of a purified state $|W\rangle$. Following a similar derivation, the counterpart of Eq. (\ref{ThV4}) is
\begin{align}\label{ThV42}
\theta_{TB}&=\mi\int_0^\tau\dif t\bigg\{-\frac{1}{2Z(t)}\sum_n\me^{-\beta E_n(t)}\left(\frac{\dot{Z}(t)}{Z(t)}+\beta \dot{E}_n(t)\right)\notag\\
&+\frac{1}{Z(t)}\sum_n\me^{-\beta E_n(t)}\Big[\langle n(t)|\frac{\dif }{\dif t}|n(t)\rangle+\langle\tilde{n}(t)|\frac{\overleftarrow{\dif} }{\dif t}|\tilde{n}(t)\rangle\notag\\&+\langle\tilde{n}(t)|\dot{\tilde{U}}(t)\tilde{U}^\dagger(t)|\tilde{n}(t)\rangle\Big]\bigg\}.
\end{align}
Since $|\tilde{n}(t)\rangle$ is simply a copy of $|n(t)\rangle$ and $\langle\tilde{n}(t)|  \overleftarrow{\frac{\dif }{\dif t}} |\tilde{n}(t)\rangle\equiv\left(\frac{\dif}{\dif t}\langle\tilde{n}(t)|  \right) |\tilde{\tilde{n}}(t)\rangle=-\langle \tilde{n}(t)|  \frac{\dif }{\dif t} |\tilde{n}(t)\rangle$, then the two terms on the second line cancel each other just as those on the first line. We finally get the thermal Berry phase of a purified state:
\begin{align}\label{ThV4b3}
\theta_{TB}&=\mi\int_0^\tau\dif t\sum_n\frac{\me^{-\beta E_n(t)}}{Z(t)}\langle \tilde{n}(t)|\dot{\tilde{U}}(t)\tilde{U}^\dagger(t)|\tilde{n}(t)\rangle, \nonumber \\
&=\mi\int_0^\tau\dif t\text{Tr}_t\left[\rho(t)\dot{\tilde{U}}(t)\tilde{U}^\dagger(t)\right].
\end{align}
Obviously, there is no geometrical term like (\ref{ThV4c2}) in this situation. Thus, the thermal Berry phase of purified states carries no geometrical information, which is essentially different from that of a thermal vacuum. Therefore, the partial transposition of the ancilla does produce a nontrivial geometrical effect that can be verified by the thermal Berry phase, in addition to its role in characterizing the separability of mixed states~\cite{QIPeresPRL,QI_book}.

\subsection{Example}\label{appe}
\textit{Example} \ref{Sec4}.1:
To illustrate the geometrical role of $\theta_{TB}$ in differentiating a purified state from a thermal vacuum, we consider a two-level system described by the Hamiltonian $\hat{H}=R\mathbf{n}\cdot \boldsymbol{\sigma}$, where $\boldsymbol{\sigma}=(\sigma_1,\sigma_2,\sigma_3)^T$ are the Pauli matrices. The unit vector $\mathbf{n}$ is parameterized as $\mathbf{n}=(\sin\theta\cos\phi,\sin\theta\sin\phi,\cos\theta)^T$. Hence, the parameter space is the two sphere $S^2$. The corresponding energy levels and eigenstates are
\begin{align}\label{EL2}
E_\pm=\pm R,~ &|R_+\rangle=\left(\begin{array}{c}
\cos\frac{\theta}{2}\\ \sin\frac{\theta}{2}\me^{\mi\phi}
 \end{array}\right),~ |R_-\rangle=\left(\begin{array}{c}
\sin\frac{\theta}{2}\\ -\cos\frac{\theta}{2}\me^{\mi\phi}
 \end{array}\right).
\end{align}
Consider a  mixed state in thermal equilibrium with temperature $T$.
The density matrix $\rho$ shares the eigenvectors with $\hat{H}$, and its eigenvalues are given by $\lambda_\pm=\frac{\me^{\mp\beta R}}{Z}$, where the partition function $Z=\me^{-\beta R}+\me^{\beta R}$. The corresponding thermal vacuum can be obtained by performing a thermal transformation on the two-mode ground state: $|0_\beta\rangle=U_\beta|R_-\rangle\otimes|\tilde{R}_-\rangle$, where the explicit expression of $U_\beta$ can be found in Appendix \ref{app:TB}.
Let the system evolve along a closed curve $C(t):=(\theta(t),\phi(t))$ on $S^2$. The Berry connection for each level is
\begin{align}\label{Bp2l}
A^\pm_B=\mi\langle  R_\pm|\dif | R_\pm\rangle=-\frac{1}{2}\left(1\mp\cos\theta\right)\dif \phi.
\end{align}

\begin{figure}[th]
\centering
\includegraphics[width=3.4in,clip]
{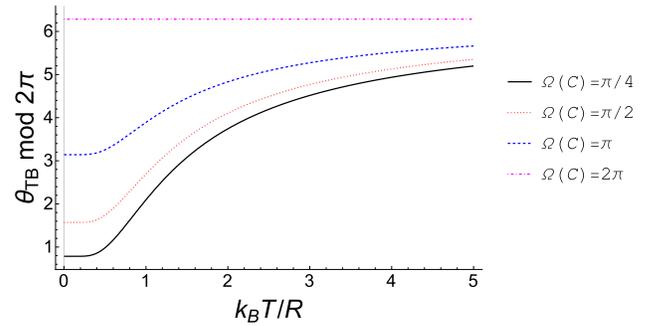}
\caption{Thermal Berry phase of the two-level system in a thermal vacuum described by Eq.~\eqref{Bp2la} as a function of temperature for different values of the solid angle in the parameter space. From top to bottom: $\Omega(C)=2\pi, \pi, \pi/2, \pi/4$.}
\label{Fig1}
\end{figure}

If a thermal vacuum of the system plus ancilla is constructed, the thermal Berry phase in this process is
 \begin{align}\label{Bp2la}
\theta_{TB}&=2\oint_C\left(\lambda_+A^+_B+\lambda_-A^-_B\right)\notag\\
&=(2\lambda_--1)\Omega(C)-4\lambda_-\pi,
\end{align}
where $\Omega(C)=\mathlarger\oint_C\left(1-\cos\theta\right)\dif \phi$ is the solid angle enclosed by the loop $C$, and $\lambda_++\lambda_-=1$ has been applied in the last step. Therefore, the thermal Berry phase in this case only comes from the contribution of the first term of Eq. (\ref{ThV4a}) and indeed carries geometrical information as the system follows a loop $C$ in the Bloch ball. If $\Omega(C)=2\pi$, i.e. $C$ is a great circle on $S^2$ such as the meridian or equator, we have $\theta_{TB}=\pm 2\pi=0$ (mod $2\pi$) at any temperature, which is a trivial phase. To visualize the result, we plot $\theta_{TB}$ vs. temperature for different values of $\Omega(C)$ in Fig. \ref{Fig1}. At zero temperature, all results reduce to twice the ordinary Berry phase of the double-mode ground state $|R_-\rangle\otimes |\tilde{R}_-\rangle$. The factor of $2$ comes from the equal contributions of both the system and ancilla.
As the infinite-temperature limit is approached, $\theta_{TB}\rightarrow 2\pi (=0$ mod $2\pi)$ for any loop $C$. In this case, the density matrix corresponds to  the center of the Bloch ball. Thus, any loop $C$ shrinks to a single point, which has a trivial topology. Hence, the thermal Berry phase is trivial in the infinite-temperature limit.
More technical details of this example, including a classification of the equivalence class of mixed states based on the Berry phase, can be found in Appendix~\ref{app:TB}.

In contrast, if a purified state of the system plus ancilla is constructed for the same case, the thermal Berry phase can be shown to be identically zero (mod $2\pi$), as it lacks geometrical information in this case. We remark that the thermal Berry phase counts the contributions from both the system and ancilla. In the case of a purified state, the total contribution vanishes at any temperature.
Therefore, this example clearly shows that the thermal Berry phase can be used to distinguish the two different representations of mixed states (except the case with infinite temperature and the case with a great circle in the parameter space).
We emphasize the result is quite general: As long as a Hamiltonian leads to a nontrivial thermal Berry phase for a thermal vacuum, the system can distinguish the result from a purified state since the thermal Berry phase is always zero for a purified state.

\section{Generalized Geometrical Phase of Mixed States}\label{Sec5}
\subsection{Generalization of parallel-transport condition}
The thermal Berry phase already predicts interesting results for differentiating a thermal vacuum and purified state. However, the physical role of the thermal Hamiltonian $\hat{H}_\beta$ is worth more investigations. Moreover, the thermal Berry phase may contain non-geometrical information. Here we presents another approach towards a geometrical generalization of the Berry phase. Note the system plus ancilla of a purified state is governed by the Hamiltonian $\hat{H}\otimes \tilde{H}$. When compared to the thermal Berry phase, we may instead consider a composite state $|W(t)\rangle$ evolving adiabatically under the composite Hamiltonian $\hat{H}\otimes \tilde{H}$, which has a more physical meaning comparing to the temperature-dependent $\hat{H}_\beta$.
Moreover, this adiabatic evolution can be incorporated
in certain composite unitary transformations that the
composite system undergoes, as
long as the related parallel-transport condition is satisfied.
By replacing $|\psi(t)\rangle$ in Eq. (\ref{PXT}) by $|W(t)\rangle$, we obtain the parallel-transport condition in this case:
\begin{align}\label{PXTm}
\langle W(t)|\frac{\dif}{\dif t}|W(t)\rangle=0,
\end{align}
where $|W(t)\rangle$ is the purified state of $\rho(t)$. The dependence of $\rho$ on $t$ is via a $t$-dependent curve $\mathbf{R}(t)$: $\rho(t)\equiv \rho(\mathbf{R}(t))$. The density matrix here describes the system while its purification includes the ancilla.
Using similar derivations, the corresponding parallel-transport conditions can also be constructed for a thermal vacuum. We present analyses of different types of unitary transformations in the following subsections.

\subsubsection{Unitary transformations involving the system and ancilla}
We consider the most general situation, where both the system and ancilla undergo unitary transformations:
\begin{align}\label{TF1}
|W(t)\rangle=U_1(t)\otimes \tilde{U}^T_2(t)|W(0)\rangle.
\end{align}
Here $U_1(0)=1=\tilde{U}_2(0)$. A previous study on a simulation of the Uhlmann phase~\cite{npj18} has implemented the latter type of transformations.
Generically, $U_1\neq \tilde{U}_2$.
The purification and density matrix respectively transform as
\begin{align}\label{TF2}
W(t)=U_1(t)W(0)\tilde{U}_2(t),\quad \rho(t)=U_1(t)\rho(0)U_1^\dagger(t).
\end{align}
Using
$|\dot{W}(t)\rangle=\left[\dot{U}_1(t)\otimes \tilde{U}^T_2(t)+U_1(t)\otimes \dot{\tilde{U}}_2^T(t)\right]|W(0)\rangle$
and the previous results, the parallel-transport condition becomes
\begin{align}\label{PXTmG}
0&=\langle W(t)|\frac{\dif}{\dif t}|W(t)\rangle  \notag\\
&=\text{Tr}_1\left[\rho(0)U_1^\dagger(t)\dot{U}_1(t)\right]+\text{Tr}_2\left[\tilde{\rho}(0)\dot{\tilde{U}}_2(t)\tilde{U}_2^\dagger(t)\right]
\end{align}
for a purified state.
Following a similar derivation, we obtain the parallel-transport condition for a thermal vacuum:
\begin{align}\label{PXTmG2}
0&=\langle 0_\beta (t)|\frac{\dif}{\dif t}|0_\beta (t)\rangle \notag \\
&=\text{Tr}_1\left[\rho(0)U_1^\dagger(t)\dot{U}_1(t)\right]+\text{Tr}_2\left[\tilde{\rho}^T(0)\dot{\tilde{U}}_2(t)\tilde{U}_2^\dagger(t)\right],
\end{align}
where $\text{Tr}A=\text{Tr}A^T$ for an arbitrary matrix $A$ has been applied for the second term.
Since the two terms in the parallel-transport condition ~\eqref{PXTmG} or ~\eqref{PXTmG2} are the integrands of the dynamical phase of the system and ancilla, both parallel-transport conditions lead to a vanishing total dynamical phase after a cycle:
\begin{align}\label{PXTmG1}
\theta_D+\tilde{\theta}_{D}=0.
\end{align}
Therefore, the dynamical phase is no longer a concern after imposing the proper condition of parallel transport for a purified state or thermal vacuum. In the following, we will study the geometrical phase accumulated in those parallel-transport processes.

\subsubsection{Unitary transformation of the system only}
By setting $\tilde{U}_2=1$, the transformation is on the system only. We will drop the subscript of $U_1$ in this subsection.
The parallel-transport condition becomes
\begin{align}\label{PXTm2}
0=\langle W(t)|\frac{\dif}{\dif t}|W(t)\rangle&=\langle W(0)|\left(U^\dagger(t)\dot{U}(t)\right)\otimes 1|W(0)\rangle\notag\\
&=\text{Tr}_1\left[\rho(0)U^\dagger(t)\dot{U}(t)\right]
\end{align}
for purified states.
Here we have applied Eq. (\ref{aO1}) to the second line, and the trace is taken over the first Hilbert space only.
Since no nontrivial transformation is imposed on the ancilla, the parallel-transport condition for the thermal vacuum is the same in this case.

Using $\rho(0)U^\dagger(t)=U^\dagger(t)\rho(t)$, we further get
\begin{align}\label{PXTm3}
\text{Tr}_1\left[\rho(t)\dot{U}(t)U^\dagger(t)\right]=0.
\end{align}
This is the parallel-transport condition (the weak version) introduced in Ref.~\cite{GPMQS1}.
Previous discussions indicate that the dynamical phase is excluded by the parallel-transport condition (see Eq. (\ref{dp1}) and its implications). It is also true here. If $U(t)$ is the time evolution operator in the first Hilbert space, then $\mi\hbar \dot{U}(t)=\hat{H}U(t)$, which implies the vanishing of the dynamical phase of the system according to Ref.~\cite{GPMQS1}:
 \begin{align}\label{PXTm4}
\theta_D=-\frac{1}{\hbar}\int_0^\tau\dif t\text{Tr}_1[\rho(t)\hat{H}]=0.
\end{align}
Hence, no dynamical phase is accumulated during parallel transport.

Since the unitary transformation preserves the norm of $|W(t)\rangle$, the condition (\ref{PXTm}) is also equivalent to
\begin{align}\label{PXTmb}
\text{Im}\langle W(t)|\frac{\dif}{\dif t}|W(t)\rangle=0.
\end{align}
The expression for a thermal vacuum has the same form.
We emphasize that either Eq. (\ref{PXTm}) or (\ref{PXTmb}) is only necessary but not sufficient for determining parallel transport. This is because $U(t)$ cannot be uniquely determined by the parallel-transport condition of mixed states, in contrast to the pure-state case. Therefore, Ref.~\cite{GPMQS1} imposed a more stringent condition by strengthening Eq. (\ref{PXTm3}) as
\begin{align}\label{PXTm5}
\langle n(t)|\rho(t)\dot{U}(t)U^\dagger(t)|n(t)\rangle=0,
\end{align}
where $|n(t)\rangle$ is an eigenvector of $\rho(t)$. This provides $N$ conditions. In general, $U$ still cannot be fully determined for mixed states since it has $N\times N$ elements. However, for a cyclic transformation satisfying $\rho(\tau)=\rho(0)$, a feasible choice is that $U(\tau)$, or even every $U(t)$, is a diagonal matrix. Then the strengthened condition (\ref{PXTm5}) is sufficient for the choice.

We emphasize that the necessary condition (\ref{PXTmb}) for parallel transport is quite loose, even a Uhlmann process with nonunitary transformations satisfies it (but not Eq. (\ref{PXTm})). Some details of the Uhlmann process can be found in Appendix \ref{appUhlmann}. Hence, Eq. (\ref{PXTmb}) can be used as a weakened parallel-transport condition in experimental simulations of the Uhlmann phase~\cite{ourPRA21}. Interestingly, different approaches for the geometrical phase of mixed states can be unified by a single condition (\ref{PXTmb}).

\subsubsection{Unitary transformation of the ancilla only}\label{appua}
We also contemplate possibilities of unitary transformations involving only the ancilla (the second Hilbert space). This can be done by setting $U_1=1$ and drop the subscript of $\tilde{U}_2$. Then,
\begin{align}\label{tmp3b}
|W(t)\rangle&=1\otimes \tilde{U}^T(t)|W(0)\rangle\notag\\
&=\sum_n\sqrt{\lambda_n(0)}|n(0)\rangle\otimes \tilde{U}^T(t)|\tilde{n}(0)\rangle
\end{align}
subject to $\tilde{U}(0)=1$. It is equivalent to
$W(t)=W(0)\tilde{U}(t)$.
Under this transformation, the density matrix of the system is unchanged: $\rho(t)=W(t)W^\dagger(t)=\rho(0)$, but a nonzero phase may be accumulated.

Applying the parallel-transport condition (\ref{PXTm}) to a purified state, we get
\begin{align}\label{PXTmII}
0=\langle W(t)|\frac{\dif}{\dif t}|W(t)\rangle&=\sum_n\lambda_n(0)\langle \tilde{n}(0)|\dot{\tilde{U}}(t)\tilde{U}^\dagger(t)|\tilde{n}(0)\rangle\notag\\
&=\text{Tr}_2[\tilde{\rho}(0)\dot{\tilde{U}}(t)\tilde{U}^\dagger(t)],
\end{align}
where the trace is over the ancilla (the second Hilbert space), and $\tilde{\rho}(0)=\sum_n\lambda_n(0)|\tilde{n}(0)\rangle\langle\tilde{n}(0)|$ is the initial density matrix. Note $\tilde{\rho}(t)=\tilde{U}(t)\tilde{\rho}(0)\tilde{U}^\dagger(t)$, i.e., the density matrix of the ancilla changes, unlike that of the system. If $\tilde{U}$ is a dynamical evolution in the ancilla space such that $\mi\hbar\dot{\tilde{U}}=\tilde{H}\tilde{U}$, then the parallel-transport condition leads to a vanishing dynamical phase of the ancilla accumulated during the process governed by the effective Hamiltonian $\tilde{H}_{\tilde{U}}$:
\begin{align}\label{PXTm4b}
\tilde{\theta}_{D}=-\frac{1}{\hbar}\int_0^\tau\dif t\text{Tr}_2[\tilde{\rho}(t)\tilde{H}_{\tilde{U}}]=0,
\end{align}
where $\tilde{H}_{\tilde{U}}=\tilde{U}\tilde{H}\tilde{U}^\dagger$.
Therefore, the extra non-geometrical information induced by Eq. (\ref{ThB1b}) of the thermal Berry phase never appears here. This approach thus leads to a more geometrical generalization of the Berry phase. 
We remark that any unitary transformation $\tilde{U}$ of the ancilla does not affect the cyclic evolution of the system because they do not contribute to the density matrix directly. Thus, even after we strengthen the parallel-transport condition $\langle \tilde{n}(0)|\tilde{\rho}(0)\dot{\tilde{U}}(t)\tilde{U}^\dagger(t)|\tilde{n}(0)\rangle=0$ like Eq. (\ref{PXTm5}), $\tilde{U}$ still cannot be fully determined. Therefore, many possibilities for this type of unitary transformations are allowed and will be illustrated by explicit examples later.

For a thermal vacuum, the parallel-transport condition implies
\begin{align}\label{PXTmI}
0&=\langle 0_\beta (t)|\frac{\dif}{\dif t}|0_\beta (t)\rangle \notag \\
&=\sum_n\lambda_n(0)\langle \tilde{n}(0)|\tilde{U}^*(t)\dot{\tilde{U}}^T(t)|\tilde{n}(0)\rangle\notag\\
&=\text{Tr}_2[\tilde{\rho}^T(0)\dot{\tilde{U}}(t)\tilde{U}^\dagger(t)],
\end{align}
Here $|0_\beta (t)\rangle$ describes thermal vacuum with the inner product of its ancilla evaluated according to Eq. (\ref{ipe3}).
Thus, if the ancilla undergoes a dynamical evolution governed by $\hat{H}^T$, $\mi\hbar\dot{\tilde{U}}=\tilde{H}^T\tilde{U}$, then the parallel-transport condition also leads to a vanishing dynamical phase of the ancilla:
 \begin{align}\label{PXTm4c}
\tilde{\theta}_{D}&=-\frac{1}{\hbar}\int_0^\tau\dif t\text{Tr}_2[\tilde{\rho}(t)\tilde{H}^T]=0.
\end{align}
Therefore, neither a purified state nor a thermal vacuum acquires a finite dynamical phase if the corresponding parallel-transport condition is followed.

\subsection{Generalized Berry phase}
A generalization of the Berry phase following the parallel-transport conditions discussed previously can now be defined as the phase accumulated when a purified state or thermal vacuum experiences cyclic parallel-transport satisfying Eq. (\ref{PXTm}). We refer to this phase as the generalized Berry phase, given by 
\begin{align}\label{Bpm}
\theta_G&=\arg\langle W(0)|W(\tau)\rangle~~ \textrm{(purified state)}, \nonumber \\
&=\arg\langle 0_\beta(0)|0_\beta(\tau)\rangle~~ \textrm{(thermal vacuum)},
\end{align}
where $t=0$ and $t=\tau$ respectively denote the initial and final parameters. For the purified state, it can be further expressed as $\theta_G=\arg\text{Tr}\left[W^\dagger(0)W(\tau)\right]$. For the thermal vacuum, it does not have such a simplification since the inner product in the ancilla space is evaluated according to Eq. (\ref{ipe3}).

When both the system and ancilla undergo a composite unitary transformation $U_1\otimes \tilde{U}^T_2$, we get
 \begin{align}\label{Bpm3II}
\theta_G=\arg\text{Tr}\left[\sqrt{\rho(0)}U_1(\tau)\sqrt{\rho(0)}\tilde{U}_2(\tau)\right]
\end{align}
for a purified state or
 \begin{align}\label{Bpm3I}
\theta_G=\arg\text{Tr}\left[\sqrt{\tilde{\rho}(0)}^TU_1(\tau)\sqrt{\tilde{\rho}(0)}^T\tilde{U}_2(\tau)\right]
\end{align}
for a thermal vacuum.
Here we have omitted the subscripts ``1'' and ``2'' because the trace can be evaluated either in the system or ancilla space. Detailed derivations are given in Appendix~\ref{app:der}. Those expressions are direct generalizations of the interferometric phase to mixed states.

The expressions simplify if the unitary transformation only acts on the system or ancilla but not both.
If only the system undergoes a unitary transformation $U$, we recover the interferometric phase~\cite{GPMQS1}
\begin{align}\label{Bpm1}
\theta_G=\arg\text{Tr}_1\left[\rho(0)U(\tau)\right]
\end{align}
for both purified state and thermal vacuum since the difference of the ancilla does not contribute.
If only the ancilla undergoes a unitary transformation $\tilde{U}$, we obtain a geometrical phase by manipulating the second Hilbert space. The result is
\begin{align}\label{Bpm2II}
\theta_G=\arg\text{Tr}_2\left[\tilde{\rho}(0)\tilde{U}(\tau)\right]
\end{align}
for the ancilla of a purified state and
\begin{align}\label{Bpm2I}
\theta_G=\arg\text{Tr}_2\left[\tilde{\rho}^T(0)\tilde{U}(\tau)\right]
\end{align}
for the ancilla of a thermal vacuum.

We consider the most generic case that both the system and ancilla undergo a cyclic adiabatic process along a closed curve $C(t):=\mathbf{R}(t)$ in the parameter space. For simplicity, we also assume $[\rho(t),\hat{H}(t)]=0$, so they share the same eigenvectors. Thus, each energy level obtains its Berry phase at the end of the process. The unitary transformations of the system and ancilla when the parameter takes the value $t$ are respectively given by
\begin{align}\label{saU}
U_1(t)&=\sum_{n}\me^{-\mathlarger{\int}_{0,C}^t\dif t'\langle n(t')|\frac{\dif}{\dif t'}|n(t')\rangle}|n(t)\rangle\langle n(0)|,\notag\\
\tilde{U}_2(t)&=\sum_{n}\me^{\mathlarger{\int}_{0,C}^t\dif t'\langle \tilde{n}(t')|\frac{\dif}{\dif t'}|\tilde{n}(t')\rangle}|\tilde{n}(0)\rangle\langle \tilde{n}(t)|,
\end{align}
where $|n(t)\rangle\equiv|n(\mathbf{R}(t))\rangle $, $|\tilde{n}(t)\rangle\equiv|\tilde{n}(\mathbf{R}(t))\rangle $ and the dynamical phase has been eliminated by the parallel-transport condition, as discussed previously. A straightforward calculation shows that the transformations satisfy
\begin{widetext}
\begin{align}\label{saU2}
\dot{U}_1(t)U_1^\dagger(t)&=\sum_{n}\left[-\langle n(t)|\frac{\dif}{\dif t}|n(t)\rangle |n(t)\rangle\langle n(t)|+\left(\frac{\dif}{\dif t}|n(t)\rangle \right)\langle n(t)|\right],\notag\\
\dot{\tilde{U}}_2(t)\tilde{U}_2^\dagger(t)&=\sum_{n}\langle\tilde{ n}(t)|\frac{\dif}{\dif t}|\tilde{n}(t)\rangle |\tilde{n}(0)\rangle\langle \tilde{n}(0)|+\sum_{nm}\me^{\mathlarger{\int}_{0,C}^t\dif t'\left(\langle \tilde{n}(t')|\frac{\dif}{\dif t'}|\tilde{n}(t')\rangle-\langle \tilde{m}(t')|\frac{\dif}{\dif t'}|\tilde{m}(t')\rangle\right)}\langle\tilde{ n}(t)|\frac{\overleftarrow{\dif}}{\dif t}|\tilde{m}(t)\rangle |\tilde{n}(0)\rangle\langle \tilde{m}(0)|,
\end{align}
\end{widetext}
which further imply
$\langle n(t)|\dot{U}_1(t)U_1^\dagger(t)|n(t)\rangle=0$ and either $\langle \tilde{n}(0)|\dot{\tilde{U}}_2(t)\tilde{U}_2^\dagger(t)|\tilde{n}(0)\rangle=0$ for a purified state or $\langle \tilde{n}(0)|\tilde{U}^*_2(t)\dot{\tilde{U}}^T_2(t)|\tilde{n}(0)\rangle=0$ for a thermal vacuum. Therefore, the condition (\ref{PXTm3}) of the system and either (\ref{PXTmII}) or (\ref{PXTmI}) of the ancilla are respectively satisfied.

At the end of a cycle ($t=\tau$), each energy level of the system acquires the corresponding Berry phase shown in Eq.~\eqref{Bp1},
and each energy level of the ancilla acquires a geometrical phase $\tilde{\theta}_{\tilde{n}}(C)=-\theta_{n}(C)$. In the space spanned by the eigenvectors of $\rho(0)$, the related unitary transformations are given by $U_1(\tau)=\tilde{U}^\dagger_2(\tau)=diag(\me^{\mi\theta_{1}},\me^{\mi\theta_{2}},\cdots)$.
Using $\sqrt{\rho(0)}=diag(\sqrt{\lambda_1(0)},\sqrt{\lambda_2(0)},\cdots)$,
Eqs. (\ref{Bpm3I}) and (\ref{Bpm3II}) both predict
\begin{equation}\theta_{G}(C)=0,
\end{equation}
so the generalized Berry phase vanishes for both purified state and thermal vacuum.
Interestingly, this happens since the geometrical contributions from the system and ancilla cancel each other exactly.

If $\tilde{U}_2=1$, the geometrical phase reduces to
\begin{align}\label{GB1}
\theta_{G}(C)=\arg\text{Tr}_1\left[\rho(0)U_1(\tau)\right]=\arg\sum_n\left[\lambda_n(0)\me^{\mi\theta_{Bn}}\right],
\end{align}
which is the argument of the weighted summation of the Berry phase factor of each level. This holds for both purified state and thermal vacuum since $\tilde{U}_2=1$ here. If $\rho$ denotes a thermal equilibrium state at temperature $T$, then
\begin{align}\label{GB2}
\theta_{G}(C)=\arg\sum_n\left[\frac{\me^{-\beta E_n(0)}}{Z(0)}\me^{\mi\theta_{Bn}}\right].
\end{align}
This also recovers the result of the interferometric phase~\cite{GPMQS1}.
We remark that the generalized Berry phase is the argument of the thermal average of the Berry phase factor if only the system experiences a unitary transformation.
This is in stark contrast with Eq. (\ref{ThV4c2}), where a weighted sum of the Berry phase is taken.

\subsection{Examples}\label{SecE}
After showing the general frameworks, we use explicit examples to demonstrate the geometrical relevance of the generalized Berry phase.
We first revisit \textit{Example} \ref{Sec4}.1 and compare the result with the thermal Berry phase. Note the Hamiltonian can be expressed as $\hat{H}=U_1R\sigma_3U^\dag_1$, where
\begin{equation}\label{U1}
U_1(\theta,\phi)=\begin{pmatrix}
\cos\frac{\theta}{2} & \sin\frac{\theta}{2} \\ \sin\frac{\theta}{2}\me^{\mi\phi} & -\cos\frac{\theta}{2}\me^{\mi\phi} \end{pmatrix}.
\end{equation}
Hence $|R_+\rangle =U_1\begin{pmatrix}
1\\ 0\end{pmatrix}$, $|R_-\rangle =U_1\begin{pmatrix}
0\\ 1\end{pmatrix}$, and $\rho(\theta,\phi)=U_1\me^{-\beta R\sigma_3}U^\dag_3$. We emphasize the unitary transformation acts on the system states only. As the system evolves along $C(t)=(\theta(t),\phi(t))$, the parallel-transport condition (\ref{PXTm5}) implies $\mi\dot{\phi}\sin^2\frac{\theta}{2}=\mi\dot{\phi}\cos^2\frac{\theta}{2}=0$, i.e. $C(t)$ can be chosen as any of the meridians. Thus the Berry phase for each level along $C(t)$ is $\theta_{B\pm}=\Omega_\pm(C)=0\mod 2\pi$ according to Eq. (\ref{Bp2l}). Thus, the generalized Berry phase $\theta_G(C)=0$ for both purified states and thermal vacua in this case. We remark on a subtlety here: If one substitutes $\Omega_\pm(C)=0$ into Eq. (\ref{Bp2la}), the thermal Berry phase $\theta_{TB}=0$ as well. However, this is because of the choice of $C(t)$: It may be chosen as any closed curve in $S^2$ for the thermal Berry phase. In contrast, it can only be the meridians for the generalized Berry phase as required by the parallel-transport condition.

The previous examples do not tell the difference between a purified state and thermal vacuum, so we present another example to show that the generalized Berry phase can distinguish them.

\textit{Example} \ref{Sec5}.1:
We consider a two-level system initially described by $\hat{H}=R\sigma_2$ at temperature $T$ with the two eigenvectors $|R_\pm\rangle$. Similarly, the ancilla space is spanned by $|\tilde{R}_\pm\rangle$. Let the composite state experience a unitary transformation acting only on the ancilla of the form $1\otimes\tilde{U}^T(t)$. Here $\tilde{U}(t)$ is an off-diagonal phase-shifting operator given by
\begin{align}\label{OFPS}
\tilde{U}(t)&=\me^{\mathlarger{\int}_{0}^t\dif t'\langle \tilde{R}_+(t')|\frac{\dif}{\dif t'}|\tilde{R}_+(t')\rangle}|\tilde{R}_-(0)\rangle\langle \tilde{R}_+(t)|\notag\\
&+\me^{\mathlarger{\int}_{0}^t\dif t'\langle \tilde{R}_-(t')|\frac{\dif}{\dif t'}|\tilde{R}_-(t')\rangle}|\tilde{R}_+(0)\rangle\langle \tilde{R}_-(t)|.
\end{align}
Here $|\tilde{R}_\pm(0)\rangle=|\tilde{R}_\pm\rangle$ and $|\tilde{R}_\pm(t)\rangle=V(t)|\tilde{R}_\pm(0)\rangle$ are generated by a unitary transformation $V(t)$. The unitarity of $\tilde{U}(t)$ can be verified via $\tilde{U}(t)\tilde{U}^\dagger(t)=|\tilde{R}_+\rangle\langle \tilde{R}_+|+|\tilde{R}_-\rangle\langle \tilde{R}_-|=1$ and $\tilde{U}^\dagger(t)\tilde{U}(t)=|\tilde{R}_+(t)\rangle\langle \tilde{R}_+(t)|+|\tilde{R}_-(t)\rangle\langle \tilde{R}_-(t)|=V(t)V^\dagger(t)=1$. Let
\begin{align}\label{OFPStmp1}
\theta_\pm(t)=\mi\int_0^t\dif t'\langle \tilde{R}_\pm(t')|\frac{\dif}{\dif t'}|\tilde{R}_\pm(t')\rangle.
\end{align}
It can be shown that
\begin{align}\label{OFPStmp2}
&\dot{\tilde{U}}(t)\tilde{U}^\dagger(t)=\me^{\mi\left(\theta_-(t)-\theta_+(t)\right)}\langle \dot{\tilde{R}}_+(t)|\tilde{R}_-(t)\rangle|\tilde{R}_-(0)\rangle\langle\tilde{R}_+(0)|\notag\\
&+\me^{\mi\left(\theta_+(t)-\theta_-(t)\right)}\langle \dot{\tilde{R}}_-(t)|\tilde{R}_+(t)\rangle|\tilde{R}_+(0)\rangle\langle\tilde{R}_-(0)|,
\end{align}
which satisfies \begin{align}\label{OFPStmp3}\langle \tilde{R}_\pm(0)|\dot{\tilde{U}}(t)\tilde{U}^\dagger(t)|\tilde{R}_\pm(0)\rangle=0.\end{align} Therefore, the parallel-transport condition holds for both the purified state and thermal vacuum. Moreover, since $\tilde{U}$ only transforms the state vector of the ancilla, the density matrix of the system remains unchanged, as we have pointed out in Sec. \ref{appua}. Thus, it is a cyclic transformation of the system as well. We further assume that $V(t)$ is induced by a closed curve $C(t)$ on the unit two-sphere $S^2$ as discussed in the example of Sec. \ref{appe}.

As a comparison, we also evaluate the thermal Berry phase first. According to Eq. (\ref{ThV3}), the thermal vacuum in this example can be written as
\begin{align}
|0_\beta\rangle=\frac{1}{\sqrt{Z}}\sum_{n=\pm}\me^{-\frac{\beta E_n}{2}}|R_n(0)\rangle\otimes \tilde{U}^T(t)|\tilde{R}_n(0)\rangle,
\end{align}
where $E_n$ and $Z$ are both invariant since the system state keeps unchanged during this evolution. For a thermal vacuum, Eqs. (\ref{ThV4a}) and (\ref{OFPStmp3}) imply $\theta_{TB}=0$. For a purified state, Eqs. (\ref{ThV42}) and (\ref{OFPStmp3}) also imply $\theta_{TB}=0$. Therefore, the thermal Berry phase cannot differentiate the purified state and thermal vacuum in this particular example. 

To evaluate the generalized Berry phase,
it can be found that
\begin{align}
\tilde{U}(\tau)=\begin{pmatrix}
0 & \me^{-\mi\Omega(C)}\\\me^{\mi\Omega(C)} & 0
\end{pmatrix}
\end{align}
at the end of the transformation,
where $\Omega(C)$ is the solid angle encircled by $C$. Note the initial density matrix of the ancilla is
\begin{align}
\tilde{\rho}(0)=\frac{1}{2}(1-\tanh\beta R\sigma_2)=\begin{pmatrix}
1 & \mi\tanh\beta R\\ -\mi\tanh\beta R & 1
\end{pmatrix}.
\end{align}
Applying Eqs. (\ref{Bpm2II}) and (\ref{Bpm2I}), the generalized Berry phase is
\begin{align}
\theta_G=\arg(-\sin\Omega\tanh\beta R)=\frac{\pi}{2}\left[1+\text{sgn}\left(\sin\Omega\right)\right]
\end{align}
for a purified state and
\begin{align}
\theta_G&=\arg(\sin\Omega\tanh\beta R)\notag\\
&=\frac{\pi}{2}\left[1-\text{sgn}\left(\sin\Omega\tanh\beta R\right)\right]
\end{align}
for a thermal vacuum.
The two expressions are off by a difference of $\pi\text{sgn}\left(\sin\Omega\tanh\beta R\right)=\pm \pi$. Thus, the generalized Berry phase indeed can distinguish the purified state and thermal vacuum at finite temperatures ($0<T<\infty$) for this setup. Therefore, the ability of the generalized Berry phase to differentiate a purified state and thermal vacuum depends on the protocol involved. There is another subtlety here associated with the infinite-temperature limit. In this case, $\tanh\beta R\rightarrow 0$, and its sign is ill-defined, causing $\theta_G$ for both purified states and thermal vacua to be ill-defined as well. However, this is not a general result at infinite temperature because $\theta_G(C)$ can be defined at any temperature, including the infinite-temperature limit, in \textit{Example} \ref{Sec4}. 1.

Due to the diversity of possible unitary transformations in the ancilla space, it is difficult to derive a set of criteria for classifying what kinds of transformations can or cannot differentiate a purified state from a thermal vacuum. Nevertheless, our examples show both possibilities for a two-level system plus its environment.

\subsection{Experimental implications}
While purified states of a two-level system incorporating environmental effects have been simulated on the IBM quantum platform~\cite{npj18}, thermal vacua of the transverse Ising model has been experimentally realized on an ion-trap quantum computer by the quantum approximate optimization algorithm~\cite{TFDPNAS20}. Moreover, partial transposition of a composite system has been approximately realized on quantum computers with various numbers of qubits~\cite{SPAPRA11,SPAPRL11,SPAPRA12}. Therefore, a comparison of the geometric effects reflected by the generalizations of the Berry phase of purified states or thermal vacua is expected to be achievable in future experiments on quantum computers or quantum simulators.

For example, one may consider two identical composite quantum systems of Example~\ref{Sec5}.1 of the generalized Berry phase and then apply a partial transposition to one of the composite systems. As a consequence, the composite system with a partial transposition corresponds to a purified state while the one without partial transposition may be viewed as a thermal vacuum. By applying parallel transport that involves the ancilla to both composite systems and extract their generalized Berry phase after a cycle, a $\pi$-phase difference is expected between the two composite systems. Given the large phase difference ($\pi$) between them after a cycle, the result is robust against small perturbations or noise from the hardware and offers another demonstration of geometrical protection of information.

We have presented two generalizations of the Berry phase, the thermal Berry phase and generalized Berry phase, for distinguishing the two state-vector representations of mixed states via the purified state and thermal vacuum. From the geometrical and physical points of view, the generalized Berry phase has more desirable properties since the thermal Berry phase is generated by a temperature-dependent thermal Hamiltonian and may carry non-geometrical information.
We caution that while the transformations can be on the system, ancilla, or both in the construction of the generalized Berry phase, an operation on the ancilla is necessary if we want to differentiate the purified state and thermal vacuum. 

\section{Conclusion}\label{Sec6}
The two state-vector representations of mixed states via purified states or thermal vacua have been developed in different branches of physics, but both have been realized on quantum computers~\cite{npj18,TFDPNAS20}. We have pointed out that their difference lies in a partial transposition of the ancilla, which has its origin in the Hilbert-Schmidt product. Available physical quantities, including previously studied geometric phases, cannot differentiate the two representations. By analogue of the adiabatic process of pure states, the thermal Berry phase has been constructed and shown to differentiate a purified state from a thermal vacuum. However, the thermal Berry phase may include non-geometrical information. The generalized Berry phase is then constructed by generalizing the parallel-transport condition to properly include the system and ancilla, and only geometrical contributions are included. Depending on the protocol and setup, the generalized Berry phase may also differentiate the purified state and thermal vacuum. Future demonstrations of the interplay between geometric effects and partial transposition of state-vector representations of mixed states on quantum computers or simulators will advance our understanding of quantum systems at finite temperatures.

\begin{acknowledgments}
H. G. was supported by the National Natural Science Foundation
of China (Grant No. 12074064).
C. C. C. was supported by the National Science Foundation under Grant No. PHY-2011360.
\end{acknowledgments}

\appendix
\section{Some Derivations}\label{app:der}
A derivation of Eq. (\ref{Bpm3II}) is outlined here:
\begin{widetext}
 \begin{align}\label{Bpm3IIa}
\theta_G&=\arg\sum_{n m}\sqrt{\lambda_n(0)\lambda_{m}(0)}\langle n(0)|U_1(\tau)|m(0)\rangle\langle \tilde{m}(0)|\tilde{U}_2(\tau)|\tilde{n}(0)\rangle\notag\\
&=\arg\sum_{nm}\langle n(0)|\sqrt{\rho(0)}U_1(\tau)\sqrt{\rho(0)}|m(0)\rangle\langle \tilde{m}(0)|\tilde{U}_2(\tau)|\tilde{n}(0)\rangle\notag\\
&=\arg\text{Tr}\left[\sqrt{\rho(0)}U_1(\tau)\sqrt{\rho(0)}\tilde{U}_2(\tau)\right].
\end{align}
\end{widetext}
where we have applied $n=\tilde{n}$ and $m=\tilde{m}$. Note there is no need to distinguish the system and ancilla since the trace in the last line can be evaluated in either space, and we accordingly omit the subscript ``1'' or ``2''. Similarly, Eq. (\ref{Bpm3I}) is evaluated as follows.
\begin{widetext}
 \begin{align}\label{Bpm3Ia}
\theta_G&=\arg\sum_{n m}\sqrt{\lambda_n(0)\lambda_{m}(0)}\langle n(0)|U_1(\tau)|m(0)\rangle\langle \tilde{n}(0)|\tilde{U}^T_2(\tau)|\tilde{m}(0)\rangle\notag\\
&=\arg\sum_{nm}\langle n(0)|U_1(\tau)|m(0)\rangle\langle \tilde{n}(0)|\sqrt{\tilde{\rho}(0)}\tilde{U}^T_2(\tau)\sqrt{\tilde{\rho}(0)}|\tilde{m}(0)\rangle\notag\\
&=\arg\text{Tr}\left[\sqrt{\tilde{\rho}(0)}^TU_1(\tau)\sqrt{\tilde{\rho}(0)}^T\tilde{U}_2(\tau)\right].
\end{align}
\end{widetext}

\section{More details of thermal Berry phase of two-level systems}\label{app:TB}
In \textit{Example} \ref{Sec4}.1, the thermal vacuum from Eq. (\ref{ThV1}) is given by
 \begin{align}\label{4.1tv}
|0_\beta\rangle&=\frac{\me^{-\frac{\beta\hat{H}}{2}}}{\sqrt{Z}}\left(|R_+\rangle\otimes|\tilde{R}_+\rangle+|R_-\rangle\otimes|\tilde{R}_-\rangle\right)\notag\\
&=\frac{1}{\sqrt{Z}}\left(\me^{-\frac{\beta R}{2}}|R_+\rangle\otimes|\tilde{R}_+\rangle+\me^{\frac{\beta R}{2}}|R_-\rangle\otimes|\tilde{R}_-\rangle\right),
\end{align}
where no extra transformation acting on the ancilla is included, in order to avoid introducing any nongeometrical information in this case. To obtain $|0_\beta\rangle$ by implementing a thermal transformation on the two-mode ground state $|R_-\rangle\otimes|\tilde{R}_-\rangle$, as indicated by Eq. (\ref{UT}), we need the knowledge of spin coherent states~\cite{Perelomov_book}. We introduce the spin operator in the system space: $\mathbf{J}=\frac{\hbar}{2}\boldsymbol{\sigma}$ that satisfies
  \begin{align}\label{J}
  &J_+\left|\frac{1}{2},-\frac{1}{2}\right\rangle=\hbar\left|\frac{1}{2},\frac{1}{2}\right\rangle,\quad J_+\left|\frac{1}{2},\frac{1}{2}\right\rangle=0,\notag\\
  &J_-\left|\frac{1}{2},-\frac{1}{2}\right\rangle=0,\quad J_-\left|\frac{1}{2},\frac{1}{2}\right\rangle=\hbar\left|\frac{1}{2},-\frac{1}{2}\right\rangle,\notag\\
  &J_z\left|\frac{1}{2},\pm\frac{1}{2}\right\rangle=\pm\frac{1}{2}\hbar\left|\frac{1}{2},\pm\frac{1}{2}\right\rangle.
\end{align}
Here $\left|\frac{1}{2},-\frac{1}{2}\right\rangle=\begin{pmatrix}
0\\ 1\end{pmatrix}$, $\left|\frac{1}{2},\frac{1}{2}\right\rangle=\begin{pmatrix}
1\\ 0\end{pmatrix}$, thus $|R_\pm\rangle=U_1\left|\frac{1}{2},\pm\frac{1}{2}\right\rangle$. Next, we introduce $\sigma_x=\begin{pmatrix}
0&1\\ 1&0\end{pmatrix}$ in the ancilla, which acts like a spin-flip operator: $\sigma_x\left|\frac{1}{2},\pm\frac{1}{2}\right\rangle=\left|\frac{1}{2},\mp\frac{1}{2}\right\rangle$. Now we construct a composite $su(2)$ algebra by
   \begin{align}
 \hat{J}_\pm=J_\pm\otimes \sigma_x,\quad \hat{J}_z=J_z\otimes 1.
\end{align}
It is straightforward to verify that $[\hat{J}_z,\hat{J}_\pm]=\pm\hbar\hat{J}_\pm$ and $[\hat{J}_+,\hat{J}_-]=2\hbar\hat{J}_z$. With the knowledge of spin coherent states~\cite{Perelomov_book}, the unitary thermal transformation $U_\beta$ can be constructed as
  \begin{align}U_\beta=U_1\otimes\tilde{U}_1\me^{\xi\frac{\hat{J}_+}{\hbar}-\bar{\xi}\frac{\hat{J}_-}{\hbar}}U^\dag_1\otimes\tilde{U}^\dag_1,\end{align}
  where $\tilde{U}_1=U_1$ and $\xi$ is a constant to be determined. Using the
Hausdorff-Campbell disentangling formula, we further have
\begin{equation}\label{De1}
\me^{\xi\frac{\hat{J}_+}{\hbar}-\bar{\xi}\frac{\hat{J}_-}{\hbar}}=\mathrm{e}^{\zeta \frac{\hat{J}_+}{\hbar}} \mathrm{e}^{\ln \left(1+|\zeta|^{2}\right) \frac{\hat{J}_z}{\hbar}} \mathrm{e}^{-\bar{\zeta} \frac{\hat{J}_-}{\hbar}},
\end{equation}
where $\zeta=\zeta(\xi)=\frac{\xi\tan|\xi|}{|\xi|}.$ Applying Eqs. (\ref{J}) and (\ref{De1}), it can be verified that
 \begin{align}\label{UT2}
& U_\beta|R_-\rangle\otimes|\tilde{R}_-\rangle=\frac{U_1\otimes\tilde{U}_1}{\sqrt{1+|\zeta|^2}}\mathrm{e}^{\zeta \frac{\hat{J}_+}{\hbar}}\left|\frac{1}{2},-\frac{1}{2}\right\rangle\otimes\left|\frac{1}{2},-\frac{1}{2}\right\rangle\notag\\
&=\frac{U_1\otimes\tilde{U}_1}{\sqrt{1+|\zeta|^2}}\left(1+\frac{\zeta}{\hbar}J_+\otimes\sigma_x\right)\left|\frac{1}{2},-\frac{1}{2}\right\rangle\otimes\left|\frac{1}{2},-\frac{1}{2}\right\rangle\notag\\
&=\frac{1}{\sqrt{1+|\zeta|^2}}|R_-\rangle\otimes|\tilde{R}_-\rangle+\frac{\zeta}{\sqrt{1+|\zeta|^2}}|R_+\rangle\otimes|\tilde{R}_+\rangle.
\end{align}
By comparing the expression with Eq.(\ref{4.1tv}), one may set $\zeta=\me^{\frac{\beta R}{2}}$ or $\xi=\arctan\me^{\frac{\beta R}{2}}$.

Next, the density matrix of the two-level system can be expressed as
\begin{align}
	\rho_{\mathbf{x}}=\frac{1}{2}\left(1-\tanh(\beta R)\mathbf{n}\cdot \boldsymbol{\sigma}\right)
\end{align}
with $\lambda_\pm=\frac{1}{2}(1\mp\tanh(\beta R))$.
Here $\mathbf{x}=-\tanh(\beta R)\mathbf{n}$ is the characteristic vector of $ \rho_{\mathbf{x}}$.
Thus, the space of all mixed states may be represented by $\mathbf{x}$ as a unit ball, the so-called Bloch ball. The pure states correspond to the boundary of the Bloch ball, which is the Bloch sphere~\cite{GPbook}. Note any density matrix physically equivalent to $\rho_{\mathbf{x}}$ can be obtained by performing a unitary transformation on $\rho_{\mathbf{x}}$: $U\rho_{\mathbf{x}}U^\dagger$. If we set $U=\frac{1}{\sqrt{2}}(1-\mi\boldsymbol{\sigma}\cdot \mathbf{a})$, it can be shown that
\begin{align}\label{dmd11b}
	U\rho_{\mathbf{x}}U^\dagger=\rho_{\mathbf{x}'}=\rho_{(\mathbf{x}\cdot\mathbf{a})\mathbf{a}+\mathbf{x}\times\mathbf{a}}.
\end{align}
Moreover, $|\mathbf{x}'|=|(\mathbf{x}\cdot\mathbf{a})\mathbf{a}+\mathbf{x}\times\mathbf{a}|=|\mathbf{x}|$. Thus, all mixed states equivalent to $\rho_{\mathbf{x}}$ form a spherical shell of radius
\begin{align}\label{r}
	|\mathbf{x}|=\lambda_+-\lambda-=\frac{\theta_{TB}+4\pi\lambda_-}{\Omega(C)}
\end{align}
inside the Bloch ball. That means all mixed states belonging to the same equivalence class have the same thermal Berry phase if the temperature and the loop in the parameter space are both specified.

\section{More discussions of Uhlmann process}\label{appUhlmann}
The discussion in the main text is focused on unitary transformations. Meanwhile, the Uhlmann process may involve nonunitary transformations~\cite{OurPRB20}. Nevertheless, the parallel-transport condition (\ref{PXTmb}) also unifies Uhlmann's parallel-transport condition~\cite{ourPRA21}. We briefly outline the arguments here. To generalize the concept of quantum holonomy to mixed states, Uhlmann
lifted the parallel-transport condition to the non-Abelian case~\cite{Uhlmann86}. Explicitly,
\begin{align}\label{PXTm6}
\dot{W}^\dagger W=W^\dagger \dot{W},
\end{align}
which is stronger than Eq. (\ref{PXTm3}) since it is a matrix equation with $N\times N$ entries.
We emphasize that this may introduce a nonunitary transformation since $\text{Tr}\rho=1$ is no longer preserved~\cite{OurPRB20}, making it totally different from the theory of interferometric phase and the formalism we discussed so far.
By taking the trace of both sides of Eq. (\ref{PXTm6}) and applying Eq. (\ref{ipe1}), we get a weaker parallel-transport condition
\begin{align}\label{PXTm7}
\langle \dot{W}(t)|W(t)\rangle=\langle W(t)|\dot{W}(t)\rangle.
\end{align}
This means if both sides are real, we return to the condition (\ref{PXTmb}).
Note the norm of $|W(t)\rangle$ is not conserved under nonunitary transformations. Consequently, the weaker condition is not equivalent to Eq. (\ref{PXTm}).
In previous simulations of the Uhlmann phase~\cite{npj18}, it is feasible to use a composite system to realize the state $|W(t)\rangle$ instead of the matrix $W(t)$. Therefore, the weakened condition (\ref{PXTmb}) is actually applied instead of Eq. (\ref{PXTm6})~\cite{npj18,ourPRA21}.
Nevertheless, we have shown that the parallel-transport conditions for the interferometric phase, the Uhlmann phase, and the generalized Berry phase can be unified via a single identity (\ref{PXTmb}).


\bibliographystyle{apsrev}

\end{document}